# Transient x-ray absorption spectroscopy of hydrated halogen atom

Christopher G. Elles, [1] Ilya A. Shkrob, [1][*] Robert A. Crowell, [1][**]

Dohn A. Arms, [2] and Eric C. Landahl [2]

[1] *Chemistry Division, Argonne National Laboratory, 9700 S. Cass Ave, Argonne, IL 60439*

[2] *Advanced Photon Source, Argonne National Laboratory, 9700 S. Cass Ave, Argonne, IL 60439*

Time-resolved x-ray absorption spectroscopy monitors the transient species generated by one-photon detachment of an electron from aqueous bromide. Hydrated bromine atoms with a lifetime of ~17 ns were observed, nearly half of which react with excess $Br^-$ to form $Br_2^-$. The K-edge spectra of the Br atom and $Br_2^-$ anion exhibit distinctive resonant transitions that are absent for the $Br^-$ precursor. The absorption spectra indicate that the solvent shell around a $Br^0$ atom is defined primarily by hydrophobic interactions, in agreement with a Monte Carlo simulation of the solvent structure.

Time-resolved x-ray scattering [1] and x-ray absorption (XA) [2-7] measurements provide detailed information about the atomic structure around short-lived species. In this Letter, we report the transient XA spectrum of a simple open shell system, the hydrated $Br^0$ atom. Details about the hydration of halogen atoms are not well known, but XA spectroscopy is uniquely suited for probing the interaction of the atom with its aqueous environment. [2,3] The hydration of halogen atoms ($X^0$) is very different from the hydration of negatively charged halide anions ($X^-$). Whereas the anion forms strong hydrogen bonds with several water molecules in the first solvent shell, [Fig. 1(a)] hydrophobic effects dominate the solvation of neutral halogen atoms. [Fig. 1(b)] There is also evidence that halogen atoms interact directly with a single solvent molecule, leading to an ultraviolet charge transfer (CT) absorption band. [8] The CT absorption promotes an electron from the water molecule onto the halogen atom. In the case of $Cl^0$, an electron spin resonance study [9] suggests that the atom forms a two center, three electron ($\sigma^2\sigma^{*1}$) bond with the oxygen atom of a double-donating water molecule [Figs. 1(c)], analogous to the much stronger Cl-Cl bond of $Cl_2^-$. [Fig. 1(d)] Similar bonding may also occur for $Br^0$ and $I^0$ in water. [10]

Recently, Pham *et al.* [7] reported XA measurements of $I^0$ atoms following biphotonic electron detachment from aqueous iodide. They observe a prominent change of the XA spectrum at the $L_1$ and $L_3$ absorption edges, but their interpretation is complicated by the reaction of $I^0$ with excess $I^-$ to form a substantial amount of $I_2^-$ and $I_3^-$ within the ~80 ps duration of the x-ray pulse. Rapid formation of molecular anions in their experiment is a consequence of the high concentration of $I^-$ (500 mmol/dm$^3$) that is necessary to overcome inefficient two-photon absorption. In contrast, efficient one-photon electron detachment from Br$^-$ at 200 nm [11] and strong XA at the K-edge of bromine allow us to use dilute aqueous solutions in which the reaction of Br$^0$ with excess anions is much slower.

The present experiment uses the laser pump – x-ray probe capabilities of beamline 7ID of the Advanced Photon Source. Fourth harmonic generation of the output from an amplified Ti:sapphire laser provides 5 µJ pulses of 200 nm light with a repetition rate of 1 kHz. A MgF$_2$ lens focuses the ultraviolet light to a diameter of 95 µm at the sample, where it crosses the x-ray beam at an angle of 4°. The laser is synchronized with the synchrotron by an active feedback control loop that adjusts the laser oscillator cavity length and the relative delay between laser and x-ray pulses is controlled electronically. The synchrotron provides tunable x-ray pulses with a duration of ~80 ps and a repetition rate of 6.54 MHz (24 bunch mode). X-rays from the undulator pass through a tunable diamond monochromator ($\Delta E/E = 5 \times 10^{-5}$) before a Kirkpatrick-Baez mirror pair focuses them to a spot size of about 25 µm in the sample. The sample is a 100 µm thick liquid jet of 5-10 mmol/dm$^3$ NaBr solution with the flat surface of the jet rotated 45° relative to the x-ray beam. Se filters absorb elastically scattered x-rays and a gated avalanche photodiode detector on each side of the jet monitors the bromine $K_\alpha$ fluorescence. We record fluorescence count rates for the two x-ray pulses immediately following the laser pulse ($\Delta t$, $\Delta t + 153$ ns) and normalize the signal to account for variations of the x-ray flux. Scanning the incident energy gives the XA spectrum at each delay time.

Fig. 2(a) compares the static XA spectrum of the bromide solution with the transient spectrum at 1 ns delay. The conversion of a fraction of Br$^-$ anions to neutral Br$^0$ atoms by the 200 nm laser pulse is evident from the resonant $1s$-$4p$ transition below the bromine K-edge. Subtracting the static spectrum of Br$^-$ from the transient spectra at delays of 1 and 154 ns gives the difference spectra $\mu_{on} - \mu_{off}$ in Fig. 2(b). The spectrum is different at the two delay times because nearly half of the $Br^0$ atoms at 1 ns react with excess $Br^-$ to form $Br_2^-$ by 154 ns, while most of the other $Br^0$ atoms recombine with the hydrated electron. Other than $Br^-$, the predominant species at 1 ns and 154 ns are $Br^0$ and $Br_2^-$, respectively. [12] The resonant transition in $Br_2^-$ is 1.6 eV higher in energy than the transition in $Br^0$, [Fig. 2] where the difference is largely due to splitting of the bonding ($\sigma_g$) and antibonding ($\sigma_u^*$) molecular orbitals relative to the atomic 4*p* orbital. [Fig 1(d)] Excitation of $Br^0$ promotes a 1*s* electron into the 4*p* vacancy produced by detaching an electron from $Br^-$, whereas the resonant transition for the $Br_2^-$ anion excites an electron to the $\sigma_u^*$ antibonding orbital. The optical transition energies for $\pi_g$-$\sigma_u^*$ and $\sigma_g$-$\sigma_u^*$ excitation are 1.7 and 3.4 eV, respectively, [13] confirming that the 1.6 eV shift of the XA band comes predominantly from the splitting of the molecular orbitals rather than a shift of the relative 1*s* orbital energy from $Br^0$ to $Br_2^-$.

The 1.6 eV spectral shift of the resonant XA band allows us to observe the reaction kinetics. Fig. 3 shows the time-dependent change in absorption at 13.473 and 13.476 keV. Predominantly $Br^0$ atoms absorb at the lower energy, therefore the decay of the absorbance indicates the loss of $Br^0$ atoms as they recombine with hydrated electrons and react with $Br^-$. Although $Br^0$ also contributes to the 13.476 keV absorption at short delay times, that signal increases on the timescale of the $Br^0$ decay due to production of $Br_2^-$. Details of the kinetic scheme and calculations of the conversion yields are given as supplemental material. [12] The kinetics give accurate estimates of the product concentrations and thus allow us to reconstruct the spectra of the transient species by subtracting the contribution from $Br^-$. These spectra are shown in Fig. 2(c), along with the spectrum of $Br^-$. The K edge absorption energy is ~5 eV higher for the transient species than for bromide. The higher energy for $Br^0$ reflects the electrostatic attraction of the outgoing electron to the positively charged core. A similar shift of ~5 eV for the absorption edge of $Br_2^-$ is somewhat surprising given the negative charge of the diatomic anion and likely reflects solvent screening and delocalization of the valence electron.

The large reduction in the modulation depth for the $Br^0$ atom relative to the $Br^-$ anion is perhaps the most intriguing feature of the recovered XA spectrum. A Monte Carlo (MC) simulation of the solvent structure provides helpful insight to understand the difference in the XA fine structure (XAFS) above the K edge. The simulation includes 200 SPC/Flex water molecules [14] and a single $Br^0$ or $Br^-$ in a supercell, with solute-water interaction potentials from refs. [15] and [10], respectively. An ensemble of 1500 snapshots at 298 K were taken from $3 \times 10^7$ MC steps to give the Br-O radial distribution functions (RDF), $g_{Br-O}(r)$, in Fig. 4(a). (The proton contribution to the XAFS spectrum is negligible.) For $Br^-$, the narrow peak at 3.2 Å is due to strong hydrogen bonding ($Br^- \cdots H$-OH) between the anion and ~6 water molecules in the first solvent shell. [15,16] The RDF for the $Br^0$ atom lacks this feature because the hydrophobic atom interacts weakly with the solvent. Instead, the atom occupies a nearly spherical cavity formed by 10-12 water molecules that are hydrogen bonded to other water molecules in the first and second solvent shells. The only distinctive feature in this RDF is a shoulder at 3 Å that corresponds to a weak $Br^0 \cdots OH_2$ adduct involving a single water molecule. For other water molecules, the $Br^0$-O distances are significantly longer, ~3.7 Å.

For further insight, we simulate the XAFS spectra of $Br^-$ and $Br^0$ [Fig. 4(b)] using the program FEFF8 [17] and nuclear configurations of water molecules with $r(Br-O) < 8$ Å from the MC ensemble. The highly organized hydrogen bonding structure of hydrated $Br^-$ gives deep oscillations in the static spectrum, but the magnitude of the oscillations is ~10 times smaller for hydrated $Br^0$ atoms. Although the calculation pertains to higher XA energies than the present experiment covers, the modulation is clearly much weaker in the reconstructed spectrum of $Br^0$ than in the $Br^-$ spectrum. [Fig. 2(c)] The contribution from the $Br^0 \cdots OH_2$ complex is small compared with the contributions from the other 10-12 O atoms in the first solvent shell and therefore does not give a strong XAFS signal.

Pham *et al.* [7] suggest that significant CT from a water molecule suppresses the $2s$-$5p$ resonance in the $L_1$ spectrum of $I^0$ by as much as 70%. No such suppression of the $1s$-$4p$ resonance is evident from our K-edge spectrum of $Br^0$. To estimate the degree of CT in the $Br^0$-water adduct, water clusters with $r(Br^0$-O$) < 5.5$ Å were extracted from the MC simulation. Water molecules outside of the extracted cluster were replaced by fractional point charges and Hartee-Fock calculations of the "embedded" clusters using a 6-311++G** basis set give a Mulliken charge of -(0.07-0.08) on the $Br^0$ atom, which would suppress the pre-edge feature very little.

In conclusion, we report the transient XA spectrum of the short-lived hydrated $Br^0$ atom and monitor its reaction to form $Br_2^-$. The solvent shell around $Br^0$ is defined primarily by weak hydrophobic interactions, with the atom residing at the center of a large "bubble." A single solvent molecule directly interacts with the atom to form a $\sigma^2\sigma^{*1}$ bond, but its XAS signature is obscured by 10-12 unbound water molecules. Calculations indicate that CT in the ground state of the Br-water adduct is too weak to suppress the $1s$-$4p$ resonance. We thank S. Ross for help with APD detectors, E. Dufresne and D. Walko for beamline assistance, L. Young and her colleagues for use of their instruments, and P. Jungwirth, C. Bressler, R. Saykally, and P. D'Angelo for helpful discussions. Use of the Advanced Photon Source was supported by the U.S. Department of Energy, Office of Basic Energy Sciences, under Contract No. DE-AC02-06CH11357.

---

\* shkrob@anl.gov, \*\* rob_crowell@anl.gov

[1]    A. Plech *et al.*, Phys. Rev. Lett. 92, 125505 (2004)

[2]    C. Bressler *et al.,* J. Chem. Phys. 116, 2955 (2002)

[3]    C. Bressler and M. Chergui, Chem. Rev. 104, 1781 (2004)

[4]    L. X. Chen, Angew. Chem. Intl. Ed. 43, 2886 (2004)

[5]    W. Gawelda *et al.*, Phys. Rev. Lett. 98, 057401 (2007)

[6]    T. Lee *et al.*, J. Chem. Phys. 122, 1 (2005)

[7]    V.-T. Pham *et al.*, J. Am. Chem. Soc. 129, 1530 (2007)

[8]    A. Treinin and E. Hayon, J. Amer. Chem. Soc. 97, 1716 (1975).

[9]    M. D. Sevilla *et al.,* J. Phys. Chem. A 101, 2910 (1997)

[10]   M. Roeselova, U. Kandor, and P. Jungwirth, J. Phys. Chem. A 104, 6523 (2000)

[11]   R. Lian *et al.*, J. Phys. Chem. A 110, 9071 (2006).

[12]   See EPAPS Document No. \*\*\*\*\*\*\*\*\*\*\*\*\*\*\*\*\* for supplementary information. This document can be reached via a direct link in the online article's HTML reference section or via the EPAPS homepage (http://www.aip.org/pubservs/epaps.html).

[13]   D. Zehavi and J. Rabani, J. Phys.Chem. 76, 312 (1972).

[14]   Y. Wu, H. L. Tepper, and G. A. Voth, J. Chem. Phys. 124, 024503 (2006).

[15]  P. D'Angelo *et al.*, J. Chem. Phys. 100, 985 (1994)

[16]  P. J. Merkling *et al.,* J. Chem. Phys. 119, 6647 (2003).

[17]  A.L. Ankudinov *et al.*, Phys. Rev. B 58, 7565 (1998).

**Figure captions.**

**FIG 1.** Sketches of the solvation structure around (a) $Br^-$ and (b) $Br^0$. Panels (c) and (d) show the schematic orbital diagrams for $Br^0 \cdots OH_2$ complex and $Br_2^-$, respectively

**FIG 2 (color online).** (a) Laser-on and laser-off x-ray absorption spectra from aqueous $Br^-$, for $\Delta t = 1$ ns. The inset shows the static spectrum of $Br^-$. (b) The difference spectra $\mu_{on} - \mu_{off}$ at delay times of 1 ns (open circles) and 154 ns (filled squares). (c) Reconstructed XA spectra of $Br^0$ (open circles) and $Br_2^-$ (filled squares). The solid line is the absorption spectrum of $Br^-$. Vertical bars indicate 95% confidence limits.

**FIG 3.** Transient XA difference $\mu_{on} - \mu_{off}$ signals for 9.8 mmol/dm$^3$ solution of NaBr observed at 13.473 keV (open circles) and 13.476 keV (filled squares). The lines are to guide the eye.

**FIG 4 (color online).** (a) Calculated Br-O RDF for hydrated $Br^-$ and $Br^0$. Oxygen coordination numbers $N$ are indicated by arrows. The inset shows the singly occupied orbital of the bromine atom inside the "embedded" water cluster from our *ab initio* calculations. (b) The simulated XAFS spectra for hydrated $Br^-$ and $Br^0$. The first oscillation corresponds to the energy range of Fig. 2(c).

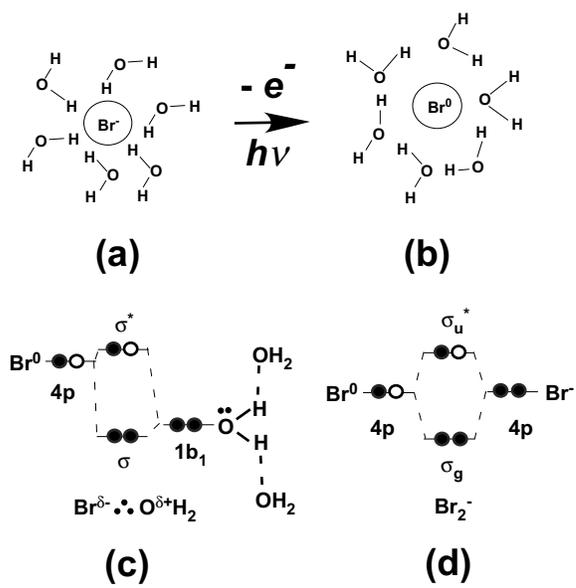

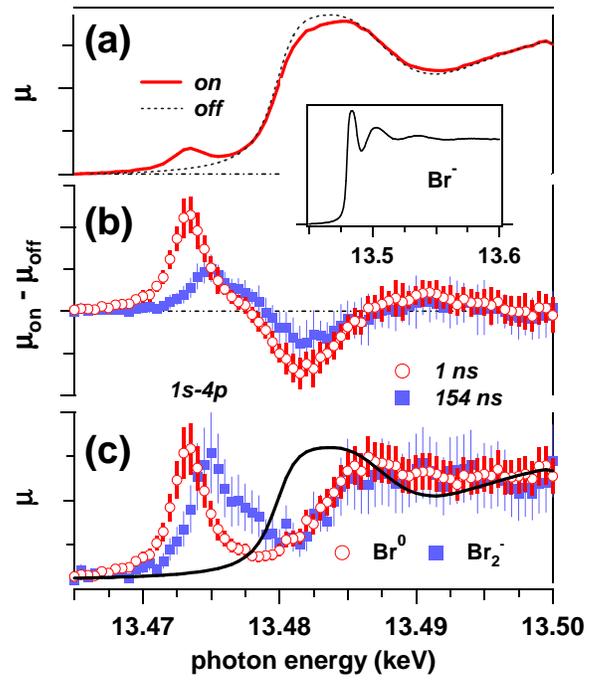

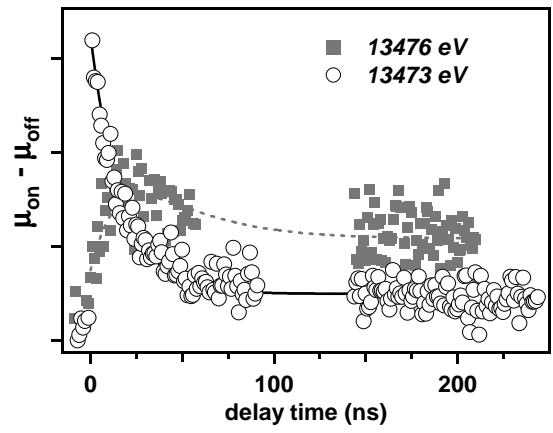

FIG 4, Elles et al.

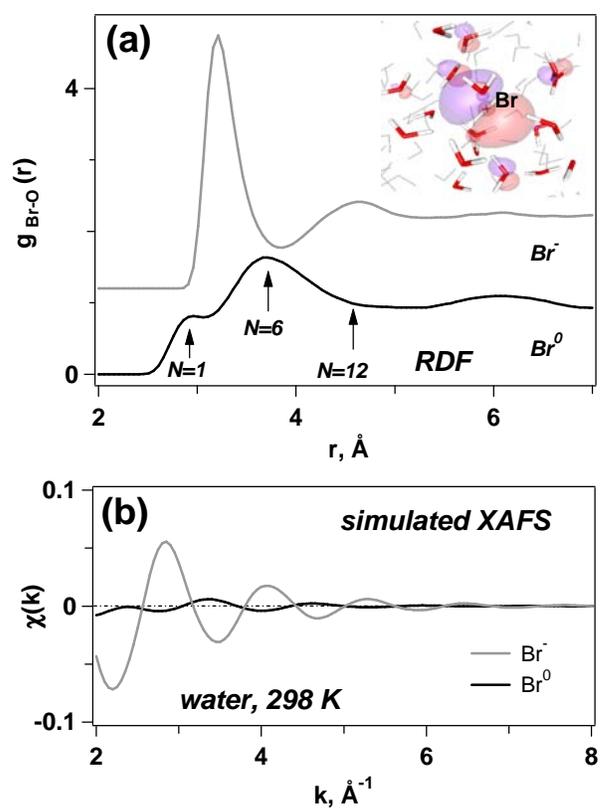

# Supplemental Information

## Transient x-ray absorption spectroscopy of hydrated halogen atom

C. G. Elles,[1] I. A. Shkrob,[1] R. A. Crowell,[1] D. A. Arms,[2] E. C. Landahl,[2]

[1] *Chemistry Division, Argonne National Laboratory, 9700 S. Cass Ave, Argonne, IL 60439;*

[2] *Advanced Photon Source, Argonne National Laboratory, 9700 S. Cass Ave, Argonne, IL 60439*

e-mail: shkrob@anl.gov, rob_crowell@anl.gov

**Reaction scheme 1.** (1 M = 1 mol/dm$^3$ = 6x10$^{20}$ cm$^{-3}$)

$$Br^- \xrightarrow{200\,nm,\,\Phi=1} (Br, e_{aq}^-) \xrightarrow{16\,ps} Br + e_{aq}^- \;(\Phi = 0.27), \quad^{1,2} \tag{1}$$

$$e_{aq}^- + Br \longrightarrow Br^- \qquad (k_2=1.4\times10^{10}\,M^{-1}\,s^{-1})\;^1 \tag{2}$$

$$Br + Br^- \longrightarrow Br_2^- \qquad (k_3=10^{10}\,M^{-1}\,s^{-1};\,K_{eq}=4.6\times10^4\,M^{-1})\;^{3,4} \tag{3}$$

$$Br_2^- + Br_2^- \longrightarrow Br^- + Br_3^- \qquad (k_4=1.7\times10^9\,M^{-1}\,s^{-1})\;^{5,6} \tag{4}$$

$$Br + Br \longrightarrow Br_2 \qquad (k_5=3\times10^{10}\,M^{-1}\,s^{-1})\;^2 \tag{5}$$

$$Br + Br_2^- \longrightarrow Br_3^- \qquad (k_6=4.6\times10^9\,M^{-1}\,s^{-1})\;^4 \tag{6}$$

$$Br_2 + Br^- \longrightarrow Br_3^- \qquad (k_7=1.5\times10^9\,M^{-1}\,s^{-1};\,K_{eq}=16.1\,M^{-1})\;^7 \tag{7}$$

$$e_{aq}^- + Br_2^- \longrightarrow 2Br^- \qquad (k_8=1.3\times10^{10}\,M^{-1}\,s^{-1})\;^5 \tag{8}$$

$$e_{aq}^- + e_{aq}^- \longrightarrow H_2 + 2HO^- \qquad (k_9=6.4\times10^9\,M^{-1}\,s^{-1})\;^8 \tag{9}$$

$$e_{aq}^- + Br_2 \longrightarrow Br_2^- \qquad (k_{10}=5.3\times10^{10}\,M^{-1}\,s^{-1})\;^9 \tag{10}$$

$$e_{aq}^- + Br_3^- \longrightarrow Br^- + Br_2^- \qquad (k_{11}=2.7\times10^{10}\,M^{-1}\,s^{-1})\;^9 \tag{11}$$

This reaction scheme was used to calculate time dependent concentrations of the transient species shown in Figure 1S and other analyses. The only variable input parameter is the initial photoconversion of the Br$^-$ to free Br$^0$ atoms. For iodide, reaction (7) has equilibrium constant of 750 M$^{-1}$, and the formation of I$_3^-$ is a major complication. For bromide, the yield of Br$_3^-$ is insignificant, as the equilibrium is shifted towards Br$_2$ at the low concentration of the bromide (< 10 mM). The yield of Br$_2$ is much lower than the yield of the Br$_2^-$ due to the occurrence of rapid reaction (10).

For 4.9 mM Br$^-$ solution (conditions for the data shown in Figure 2), the estimated ratio [Br$_2$]/[Br$_2^-$]≈0.13 at the delay time of 154 ns (that approximately corresponds to the maximum yield of the Br$_2^-$) and [Br$^0$]≈0, so the spectrum is dominated by the Br$_2^-$. At the delay time of 1 ns, [Br$_2$]/[Br$^0$]≈0.06 and [Br$_2^-$]/[Br$^0$]≈0.09, and the spectrum is mainly from the Br$^0$.

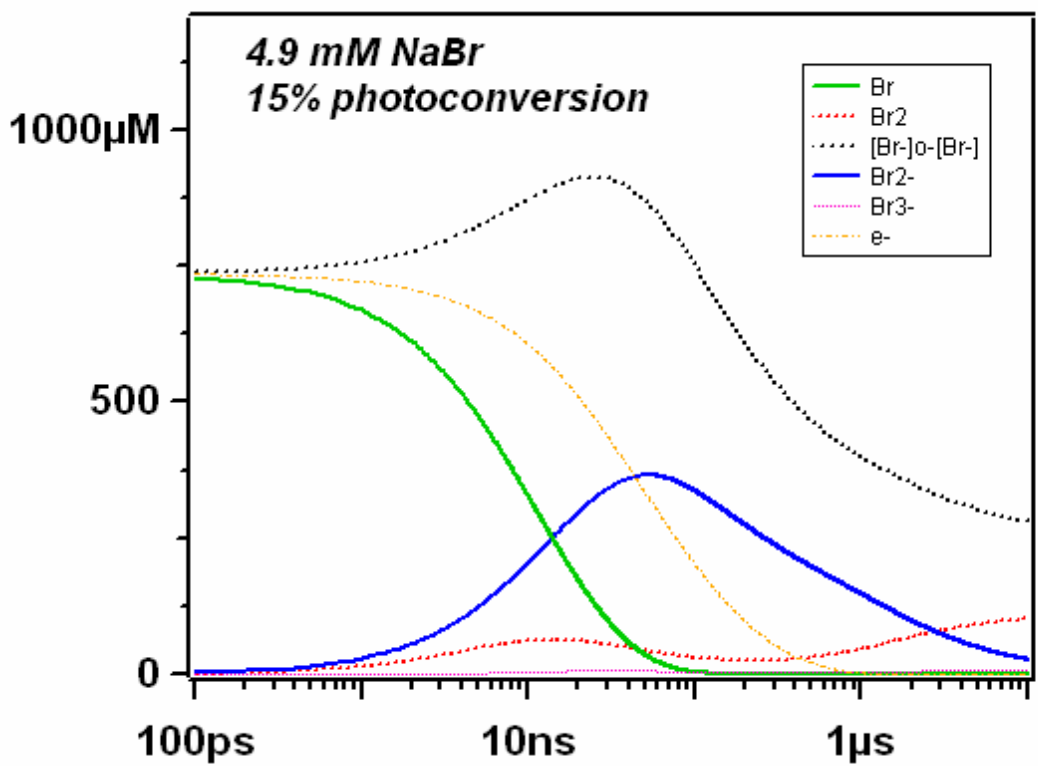

**FIG 1S.** Simulated kinetics of bromine containing species in an aqueous solution, for 15% initial photoconversion of 4.9 mM Br$^-$ to free Br$^0$ atoms.

The decay kinetics of the Br$^0$ are mainly via reactions (2) and (3). Due to the occurrence of bimolecular reaction (2) in the bulk, the apparent rate constant $k_{-Br}$ of the Br$^0$ atom decay increases as [Br$^0$] increases. For 9.8 mM Br$^-$ solution, we obtained $(6.2\pm0.4)\times10^7$ M$^{-1}$ s$^{-1}$ (Figure 3) which is significantly greater than $k_3$[Br$^-$]. Using the calculated $k_{-Br}$ vs. photoconversion efficiency as a calibration plot allowed us to estimate the photoconversion as 12-15%. For this photoconversion, the calculated ratio of [Br$_2^-$] at 154 ns and [Br$^0$] at 1 ns is 0.45-0.48, which is close to the experimental ratio of the *1s-4p* peak amplitudes (0.43-0.47). Direct calculation of the photoconversion using the extrapolated quantum yield at 1 ns (≈0.29) [1] and the extinction coefficient of the Br$^-$ at 200 nm (10$^4$ M$^{-1}$ cm$^{-1}$) that included depletion of the laser light along the optical path and laser beam refraction in the sample (*n*=1.4) gave the estimate of 13% at [Br$^-$]=9.8 mM for the 200 nm laser power of 4.7 µJ.

**References for the reaction scheme:**

1. The rate constant $k_2$ in the water bulk was estimated from picosecond geminate recombination kinetics obtained for the hydrated electron ($e_{aq}^-$) in Ref. 11 of the paper. The rate constant of reaction (2) is given by $k_2 = 4\pi R_{eff} D$, where *D* is the mutual diffusion constant and $R_{eff} = a(1-p_d)$ is the effective radius of recombination, which is the product of the Onsager radius $a$ for the potential well (that exists due to the weak attraction between hydrated $Br^0$ and $e_{aq}^-$) and $(1-p_d)$, which the probability of recombination. The Onsager radius can be estimated from the dimensionless parameter $\alpha = ap_d\sqrt{W/D}$ that was estimated to be 0.57 in Ref. 11 by fitting the geminate recombination kinetics observed for the hydrated electron; the rate constant $W$ in the expression for this parameter is the reciprocal residence time of the geminate pair inside the potential well (that was estimated to be ca. 16.5 ps in Ref. 11). For the reported $D$=4.5x10$^{-5}$ cm$^2$/s, $a$=5.8 Å, and $p_d$=0.27, we obtain $R_{eff}$=4.2 Å and $k$=1.4x10$^4$ M$^{-1}$s$^{-1}$. The terminal quantum yield $\Phi$ of the $e_{aq}^-$ (attained by the completion of the geminate recombination stage) for 193 nm photoexcitation of aqueous Br$^-$ is 0.365 (as determined by nanosecond laser photolysis by M. C. Sauer *et al.* J. Phys. Chem. A 108, 5490 (2004)). The estimate of $\Phi \approx 0.27$ for 200 nm photoexcitation of the bromide was obtained in picosecond laser photolysis experiment in Ref. 11.

2. The rate constant for recombination of two aqueous I$^0$ atoms was taken from A. J. Elliot, Can. J. Chem. 70, 1658 (1992)

3. D. Zehavi and J. Rabani, J. Phys. Chem. 76, 312 (1972)

    Y. Liu *et al.,* J. Phys. Chem. A, 106, 11075 (2002).

4. J. Lind *et al.,* J. Am. Chem. Soc. 113, 4629 (1991)

5. M. S. Matheson *et al.,* J. Phys. Chem., 70, 2092 (1966)

6. B. G. Ershov *et al.,* Phys. Chem. Chem. Phys., 4, 1872 (2002)

7. M.-F. Ruasse *et al.,* J. Phys. Chem. 90, 4382 (1986).

8. K. H. Schmidt and D. M. Bartels, Chem. Phys. 190, 145 (1995).

9. H. A. Schwarz and P. S. Gill, J. Phys. Chem. 81, 22 (1977).